\documentclass[final,5p,times,twocolumn,plainnat]{elsarticle}

\usepackage{graphicx}
\usepackage{dcolumn}
\usepackage{bm}
\usepackage{comment} 
\usepackage{float}
\usepackage{bigints}
\usepackage[utf8]{inputenc}
\usepackage{esint}
\usepackage{wrapfig}
\usepackage{amsfonts}
\usepackage{epstopdf}
\usepackage[arrowdel]{physics}
\usepackage[T1]{fontenc}
\usepackage{amsmath,amssymb,amsthm}
\usepackage{tikz}
\usepackage{cancel} 
\usepackage{color}
\usepackage{subcaption}
\usepackage[shortlabels]{enumitem}
\usepackage{listings}
\usepackage{xcolor} 
\usepackage{hyperref} 
\usepackage{breqn}
\usepackage{booktabs}
\usepackage{blindtext}
\usepackage{cuted}

\definecolor{aocol}{rgb}{0.0, 0.6, 0.0}

\newcommand{\sect}[1]{\ref{sec: #1}}

\begingroup\makeatletter
\catcode`,=\active
\global\let\breqn@comma,
\protected\gdef,{\ifmmode\expandafter\breqn@comma\else\expandafter\active@comma\fi}
\endgroup

\def\orcid#1{\href{https://orcid.org/#1}{\!\includegraphics[keepaspectratio,width=0.7em]{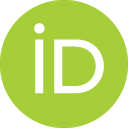}}}

\journal{Physics Letters B}

\begin{document}

\begin{frontmatter}

\title{Addressing the ground state of the deuteron by physics-informed neural networks}

\author[1]{Lorenzo Brevi \orcid{0009-0005-8135-7363}\,}
\author[1,2]{Antonio Mandarino \orcid{0000-0003-3745-5204}\,}
\author[1,3]{Carlo Barbieri \orcid{0000-0001-8658-6927}\, }
\author[1,3]{Enrico Prati \orcid{0000-0001-9839-202X}\, }
\ead{enrico.prati@unimi.it}

\affiliation[1]{organization={Dipartimento di Fisica "Aldo Pontremoli", Università degli studi di Milano},
            addressline={Via Celoria 16}, 
            city={Milano},
            postcode={20133}, 
            country={Italy}}
\affiliation[2]{organization={International Centre for Theory of Quantum Technologies, University
of Gdansk},
            addressline={Jana Ba\.zy\'nskiego 1A}, 
            city={Gda\'nsk},
            postcode={80-309}, 
            country={Poland}}
\affiliation[3]{organization={INFN, Sezione di Milano},
            addressline={Via Celoria 16}, 
            city={Milano},
            postcode={20133}, 
            country={Italy}}

\date{\today}

\begin{abstract}

Machine learning techniques have proven to be effective in addressing the structure of atomic nuclei. Physics$-$Informed Neural Networks (PINNs) are a promising machine learning technique suitable for solving integro-differential problems such as
the many-body Schr\"odinger problem.
So far, there has been no demonstration of extracting nuclear eigenstates 
using such method. Here, we tackle realistic nucleon-nucleon interaction in momentum space, including models with strong high-momentum correlations, and demonstrate highly accurate results for the deuteron. 
We further provide additional benchmarks in coordinate space.
We introduce an expression for the variational energy that enters the loss function, which can be evaluated efficiently within the PINNs framework.
Results are in excellent agreement with proven numerical methods, with a relative error between the value of the predicted binding energy by the PINN and the numerical benchmark of the order of $10^{-6}$. Our approach paves the way for the exploitation of PINNs to solve more complex atomic nuclei.
\end{abstract}


\end{frontmatter}

\section{Introduction} \label{sec: Intro}

Machine learning (ML) techniques have increasingly impacted the physical sciences in recent years, in some cases being able to outperform more traditional computational methods~\cite{AlphaFold3}, while in others providing novel approaches to solve scientific problems~\cite{Noe2019Science}. The opportunity to directly solve quantum many-body systems was realized for the first time for the spin lattice~\cite{CarleoRBM2017}. 
Neural networks can be used to represent wavefunctions of complex quantum systems, by leveraging the property of exceptionally efficient universal approximators~\cite{hornik1989multilayer,maronese2022quantum,medvidovic2024neural}. 
The combination of the variational principle with machine learning optimization techniques makes it possible to find the true ground state of the system through recursive computation and minimization of the expectation value of the Hamiltonian operator.
Such approach, known as neural network quantum states (NQS), can be considered an extension of the variational Monte Carlo (VMC) method and has a similar computational cost, but it can achieve significantly higher accuracy. The NQS framework has since been enhanced through successful applications in solid-state physics~\cite{Yoshioka2021Solids} and quantum chemistry~\cite{Hermann2023RevewQC}, often reaching comparable accuracy to the computationally much more expensive diffusion Monte Carlo (DMC) method. 
Recent developments have addressed the computation of the first low-lying excited molecular states~\cite {Pfau2024Science} and have been extended to advanced deep learning architectures such as transformers~\cite{glehn2023SelfAtt}. 
NQS-based simulations are typically more complicated for nuclear physics, due to the complexity of the nuclear force among protons and neutrons~\cite{RevModPhys.81.1773,MACHLEIDT2011}. Ref.~\cite{Keeble2023deuteron} provided a first proof-of-principle for the deuteron -- the simplest nucleus formed by one neutron and one proton -- in terms of standard feed-forward networks. The research was later extended to few-nucleon systems using Jastrow inspired networks~\cite{Adams2021prl} and an adaptation of the FermiNet~\cite{Pfau2020FNet} architecture to nuclei~\cite{Yang2023FNet}. 
In general, nuclear structure theory highly benefits from innovation in \textit{ab initio} nuclear many-body methods. Some of us have recently addressed diagrammatic Monte Carlo methods~\cite{Brolli2025DiagMC} and quantum algorithms for determining the energy spectrum of nuclei~\cite{VQE_lmg, Barone_2024, nigro2025simulation}.

The NQS approach is best classified as reinforcement learning because no training or target data is provided to learn about many-body correlations. Rather, the algorithm samples directly the expectation value of the Hamiltonian and uses the variational principle as a guiding rule to constrain the wave function. Because of this feature, NQS could also be referred to as \textit{science-driven learning}. Yet, it relies on only one piece of physics information, the variational principle, to learn about the quantum many-body structure.
The latter limitation could be overcome by exploiting physics-informed neural networks (PINNs),  introduced in Ref.~\cite{RAISSI2019686} to solve partial differential equations (PDEs). The PINN framework focuses on integrating all possible physical knowledge of the problem into the loss function of the network. In most cases, it includes boundary conditions and physical symmetries as well as measured data points where the solution is partially known, making the approach particularly suited to studies of fluid dynamics~\cite{Fluids1,Fluids2,Fluids3}. However, PINNs can solve differential equations by exploiting the sole physical constraints of the system. Exploratory applications to the Schr\"odinger equation focused on one-dimensional elementary systems \cite{PINNschrod,harcombe2023physicsinformed}. More recently, some of us have employed PINNs to solve for both the ground state and excited states of the anharmonic oscillator, a non-trivial and non-integrable quantum system~\cite{Brevi2024,tutorial}.

Here we solve for the ground state of the deuteron based on the PINN framework. We introduce a new strategy to compute eigenenergies that is a natural extension of the standard PINN approach.
The improvement allows us to tackle realistic models of the nucleon-nucleon interaction in momentum space, even with strong high-momentum correlations, and obtain highly accurate benchmarks. We also provide an example of analogous simulations in coordinate space. To our knowledge, these are the first applications of PINNs to an \textit{ab initio} nuclear structure problem.

Section~\ref{sec: PInns} gives an overview of the PINN method and the techniques for training its networks.
Sec.~\ref{sec: metrics} describes the metrics used to quantify the accuracy of our solutions. Next, Secs.~\ref{sec: res_r} and~\ref{sec: Res_k} reports on the computations in coordinate and momentum space, respectively. All results are discussed in Sec.~\ref{sec: compare}.

\section{Physics-Informed neural network for the eigenvalue problem} \label{sec: PInns}

We discuss the implementation of PINNs for solving the Schr\"odinger's equation, emphasizing the novelties introduced to better converge its eigenvalues.  For a more in-depth review refer to previous work on the subject~\cite{tutorial,Brevi2024}.
A generic Physics-informed neural network is made to encode all prior knowledge about the physical system, including the differential equation itself that one aims at solving~\cite{RAISSI2019686}. The general form for its loss function is 
\begin{equation} \label{eq: loss_generic}
    \mathcal{L}_{PINN} = \mathcal{L}_{PDE} + \mathcal{L}_{phys} + \mathcal{L}_{data} \, ,
\end{equation}
where \(\mathcal{L}_{data}\) is the standard training loss of a supervised problem, given as the discrepancy between its output and training labels. This is omitted in our computations since we do not have data points.  
\(\mathcal{L}_{phys}\) encodes to the losses given by the physical constraints of the system, such as boundary conditions and initial conditions.
Lastly, \(\mathcal{L}_{PDE}\) is the mean squared error or sum of squared errors of the differential equation. For instance, consider a generic PDE written in terms of its implicit solution \(\mu(t)\) 
and the derivative \(\partial_t\mu(t)\) with the independent variable $t$:
\begin{equation} \label{eq: generic_PDE}
    \partial_t\mu(t) + N[\mu;\xi] = 0 \, ,
\end{equation}
where $N$ is a nonlinear operator and \(\xi\) is some set of parameters. Then, one can write \(\mathcal{L}_{PDE}\) for a network designed to solve this equation as:
\begin{equation} \label{eq: generic_PDE_loss}
    \mathcal{L}_{PDE}=\frac{1}{N_c}\sum_{i=1}^{N_c}(\partial_t\mu_{net}(x_i) + N[\mu_{net}(x_i);\xi])^2
\end{equation}
where \({x_i}\) is the set of $N_c$ chosen \emph{collocation points} where the PDE will be evaluated.
Partial derivatives can be computed exploiting the same automatic differentiation techniques used to train the weights of standard neural networks. Here, we also consider the derivatives of  functions of the neural network with respect to its inputs.

This approach enables the training of neural networks with noisy or very little data, and even in the absence of labeled data as in this work. The eigenfunction and the corresponding eigenvalue of the deuteron must satisfy certain constraints, including boundary conditions, the normalization of the wavefunction, and the condition that it solves the Schr\"odinger's equation. 
In addition, we impose that the solution minimizes the energy since the deuteron has no bound excitations and we are only targeting the ground state. In this work we focus on the radial Schr\"odinger equation
, but the generalization to several spatial dimensions is straightforward. 

Hence, the PINN wavefunction $\psi(x)$ is computed in a one-dimensional interval, $x\in [0,M]$, where $x$ is the radial coordinate in either position or momentum space.
The physical constraint due to boundary conditions is encoded in the loss function\footnote{In Eq.~\eqref{eq: L_BC}, and similarly for Eq.~\eqref{eq: norm_loss_deut}, the ${N_c}/2$ factor ensures that these losses remains balanced with respect to the other loss functions that include all collocation points.}

\begin{equation} \label{eq: L_BC}
    \mathcal{L}_{BCs} = \frac {N_c}2 \left( \left| \phi(x=0) \right|^2+ \left| \psi(x=M) \right|^2 \right) \,,
\end{equation}
where $\phi(x)= x \, \psi(s)$ is used to impose the correct boundary condition for partial waves of angular momentum $L=$~0. The second term in Eq.~\eqref{eq: L_BC} ensures that the wavefunction vanishes at large distances, as expected for bound states.

The normalization condition is imposed on the integral
\begin{equation} \label{eq: norm_integ}
   \mathcal{I}[\psi]  = 4\pi \int_{0}^M |\psi(x)|^2 \; x^2 \;dx
\end{equation}
and can be implemented in two different ways. First, for the spatial coordinate case of Sec.~\ref{sec: res_r} we follow  the auxiliary output method from Ref.~\cite{YUAN2022111260}. We add one additional output, $\nu(x)$, to the neural network that will be trained to reproduce the integral
\begin{equation} \label{eq: auxiliary_norm}
    \nu(x) = \int_{0}^x |\psi(x')|^2\; (x')^2 \; dx' \,,
\end{equation}
so that $\mathcal{I}[\psi] = 4\pi\,\nu(M)$. 
This new network output is trained by introducing the partial `integration' loss 
\begin{equation} \label{eq: auxiliary}
  \mathcal{L}_{int} =  \sum_{i=0}^{N_c}\Bigl(\left|\partial_x\nu(x_i) - x^2 \, |\psi(x_i)|^2 \right| + \left|\min(0,\partial_x\nu(x_i))\right|\Bigr)^2 .
\end{equation}
This allows to compute the normalization integral in a mesh-free way, but at the cost of additional computational overhead. The last term in parentheses in Eq.~\eqref{eq: auxiliary}  imposes that the output $\nu(x)$ is monotonically increasing, as expected from Eq.~\eqref{eq: auxiliary_norm}.

The second approach to the normalization consists in computing Eq.~\eqref{eq: norm_integ} directly from finite difference methods and it will be employed in Sec.~\sect{Res_k}. This results in less computationally demanding effort but, conversely, it can introduce discretization errors. The decision of which of the two methods may result more effective depends on a case-by-case basis.
Regardless of how $\mathcal{I}[\psi]$ has been computed, we obtain the normalization loss
\begin{equation}
    \label{eq: norm_loss_deut}
    \mathcal{L}_{norm}=N_c \, \Bigl( \, \left|\mathcal{I}[\psi] - \ln{\left(\mathcal{I}[\psi]\right)} - 1 \right| + \left|\nu(0)\right| \, \Bigr) \, ,
\end{equation}
where the last term in parentheses is used only with the auxiliary output and ensures that $\int_{0}^0 \norm{\psi(x)}^2\;dx = 0$.
Note that the logarithm term in Eq.~\eqref{eq: norm_loss_deut}, even if not mandatory, produces efficient minimization because it implies that $\mathcal{I}[\psi]=0$ leads to an infinite loss, while $\mathcal{I}[\psi]=1$ is a global minimum. Otherwise, $\psi(x) = 0 \; \forall x$ would be a local minimum, making it harder to train the network.

The PDE term in Eq.~\eqref{eq: loss_generic} imposes the solution
the Schr\"odinger equation, $H\ket{\psi} = E\ket{\psi}$, where $H$ is the nuclear Hamiltonian and $E$ is the eigenvalue.
The corresponding loss, $\mathcal{L}_{Schrod}$, is the one that includes the nuclear Hamiltonian we are trying to solve, thus defining the specific physical system. To implement the PDE condition
we first need to compute the expectation value of the energy with respect to the network output, $\ket{\psi}$, as
\begin{equation}
    \label{eq: E}
    E =  \frac{\langle \psi | H |\psi \rangle}{\mathcal{I}[\psi]}
     = \sum_{j=0}^n w_j \frac{\psi(x_j) \, [H\psi](x_j)}
     {\mathcal{I}[\psi]} \,,
\end{equation}
where $w_j$ are integration weights and $[H\psi](x_j)$ indicates the state $H\ket{\psi}$ evaluated at the sampling point $x_j$. 
Note that $E$ could be computed either via finite differences, Eq.~\eqref{eq: E}, or using the auxiliary output method described above. We always use Eq.~\eqref{eq: E} in this work.
The PDE term of the loss is then,
\begin{equation}
    \label{eq: loss_eq_actual}
    \mathcal{L}_{Schrod} = \sum_{i=0}^{N_c} \left\{ [H\psi](x_i)- E \psi(x_i) \right\} \,.
\end{equation}
The structure of the PINN allows a very efficiently computation of $[H\psi](x_j)$, and hence Eqs.~\eqref{eq: E} and~\eqref{eq: loss_eq_actual}, by exploiting automatic differentiation for all relevant derivatives that enter the Hamiltonian operator $H$.

Finally, the variational principle is implemented by the loss
\begin{equation}
    \label{eq:L_var}
    \mathcal{L}_{var} =  \min\left[0,\left(bE\right)^c\right] + d(t)e^{a\left(E-E_{0}\right)} \, ,
\end{equation}
where $a$, $b$ and $E_0$ are real-valued hyperparameters, the $c$
takes only odd integers values to preserve the sign of $bE$, and
$E$ represents the energy expectation value of Eq.~\eqref{eq: E}.
The second term of Eq.~\eqref{eq:L_var} is used to lead the network toward the lowest energy states. Here, the quantity $d(t) = 0.9999^t$, where $t$ enumerates the training epochs, is a decay parameter that goes to zero as the training proceeds, ensuring that the variational principle improves the early stages of training without damaging the latter stages due to the lack of a minimum of this term when $E$ assumes the correct value.
In general, Eq.~\eqref{eq:L_var} has the purpose to both improve the early stages of training and to facilitate convergence toward the ground state. 
Note that Eq.~\eqref{eq:L_var} also works in the case of excited states, if any.

With all the above definitions, the overall PINN loss function can be summarized as
\begin{align}
\label{eq: loss_overall}
    \mathcal{L}_{PINN} ={}& \alpha_{Sch}\,\mathcal{L}_{Schrod} + \alpha_{int}\,\mathcal{L}_{int} + \alpha_{BCs}\,\mathcal{L}_{BCs} + \nonumber \\
    & \alpha_{norm}\,\mathcal{L}_{norm} + \alpha_{var}\,\mathcal{L}_{var} \,,
\end{align}
where we have added appropriate weights to each term.

\subsection{Evaluation metrics} \label{sec: metrics}
We now turn to the two evaluation metrics utilized to benchmark the quality of the results. Being the approach not supervised, these are not used for training they but reveal interesting insights.
The main quantity is the relative error of the predicted energy with respect to the exact numerical solution. We choose a signed metric to indicate whether the network overestimates or underestimates the energy:
\begin{equation}\label{eq: rel_error_e_numerical}
    err_{E,Num} = \frac{E_{Num} - E_{PINN}}{E_{Num}} \,. 
\end{equation}

The second indicator concerns the eigenvector and it is the fidelity between the predicted wave function and the exact one \cite{fid1, fid2, fid3},

$\mathcal{F}_\psi = |\braket{\psi_{Num}}{\psi_{PINN}}{}|^2$.

We evaluate this numerically as
 \begin{equation}\label{eq: sim_wf_discrete}
     \mathcal{F}_\psi = \left|\sum_{i=0}^{N_c} \psi_{PINN}(x_i) \, \overline{\psi_{Num}(x_i)} \right|^2 \, ,
 \end{equation}
where the {\(x_i\)} are a set of points spanning the whole domain, and \(\psi_{Ex}(x)\) and \(\psi_{PINN}(x)\) are wave functions from the exact numerical solution and obtained from the PINN network, respectively. Both of the vectors are properly normalized before evaluating Eq.~\eqref{eq: sim_wf_discrete}.
In this work, we chose to compute $\mathcal{F}_\psi$ on the same $N_c$ collocation points used for training.

\section{Results} \label{sec: results}

We now turn to the results for the deuteron ground state within the PINNs framework. We first focus on the Minnesota potential in coordinate space from Ref.~\cite{THOMPSON197753}. This is a particularly simplified model for the nuclear force but one that allows us to test the PINN approach on the integro-differential form for the the Schr\"odinger equation, involving spatial derivatives for the kinetic energy term. We then turn to momentum space and consider two different realistic models of the nuclear interaction. The first is the so-called next-to-next-to-next-to-next lowest order ($N{}^4LO$) interaction which is at fifth-order in chiral effective field theory ($\chi$EFT) expansion.
The $\chi$EFT is currently the best performing paradigm for the derivation of nuclear Hamiltonians~\cite{MACHLEIDT2011, RevModPhys.81.1773, Machleidt_2016, RevModPhys.92.025004, doi:10.1142/S0218301317300053, MACHLEIDT2024104117} and it is exploited in most state-of-the-art ab initio simulation of nuclei~\cite{Coraggio2020FrontBook}. Here, we exploit the $N^4LO$ two-nucleon force from Ref.~\cite{Entem2017_EMNpots_PRC} with a cutoff on the relative momentum of 550~MeV/c.
The second interaction is the CD-Bonn model from Ref.~\cite{cdbonn}. This is also a high-precision potential but built to induce strong short-range correlations---hence, high momenta in the wave function---and it is therefore the most challenging interaction of the three to diagonalize. We focus on the (one-dimensional) radial Schr\"odinger equation for the deuteron and deal with $S$ ($L$=0) and $D$ ($L$=2) partial waves. The coupling among angular momentum and spin-isospin already implies the fulfillment of the Pauli principle and no loss function for constraining antisymmetry is required.
In a more general framework with more nucleons, the Pauli principle can be effectively encoded into the network architecture through determinant wave functions, see for example Refs.~\cite{nn_quantum_states,Pfau2020FNet}, or partially enforced using additional dedicated losses.

\subsection{Deuteron in position space} \label{sec: res_r}

\begin{table}[t] 
    \centering
    \begin{tabular}{l|c|l|c}
    \hline
     Loss ($\mathcal{L}_i$)& Weight ($\alpha_i$) &  Loss ($\mathcal{L}_i$)& Weight ($\alpha_i$)  \\
     \hline\hline
      $\mathcal{L}_{Schrod}$ & $10^{-4}$ &    $\mathcal{L}_{norm}$ & $1$ \\
      \hline
      $\mathcal{L}_{int}$ & $10^{-3}$ &  $\mathcal{L}_{var}$ & $10^6$\\
      \hline
      $\mathcal{L}_{BCs}$ & $10^2$ & \\ 
      \hline\hline
    \end{tabular}
    \caption{Optimal weights for the different partial losses entering Eq.~\eqref{eq: loss_overall}, adopted to compensate for the different scales of each term.}
    \label{tab: weights_r}
\end{table}

We implemented the PINNs using a feed-forward network with $6$ hidden layers of $256$ neurons each. The network was trained using the overall loss from Eq.~\eqref{eq: loss_overall} computed over $4097$ collocation points and with the weights from Table~\ref{tab: weights_r}.  It is important to note that the different partial losses can have very different scales, and this is taken into account by the choice of the weights.
For the variational loss hyperparameters, we used $E_{0} = -2$~MeV, $a = 0.8$~MeV$^{-1}$, $b = 10^4$~MeV$^{-1}$ and $c = 1$.  The main metric used to evaluate the network's performance is the accuracy with respect to the energy obtained from the exact diagonalization of the Hamiltonian.

\begin{figure}[h]
    \centering
    \includegraphics[width=0.7\linewidth]{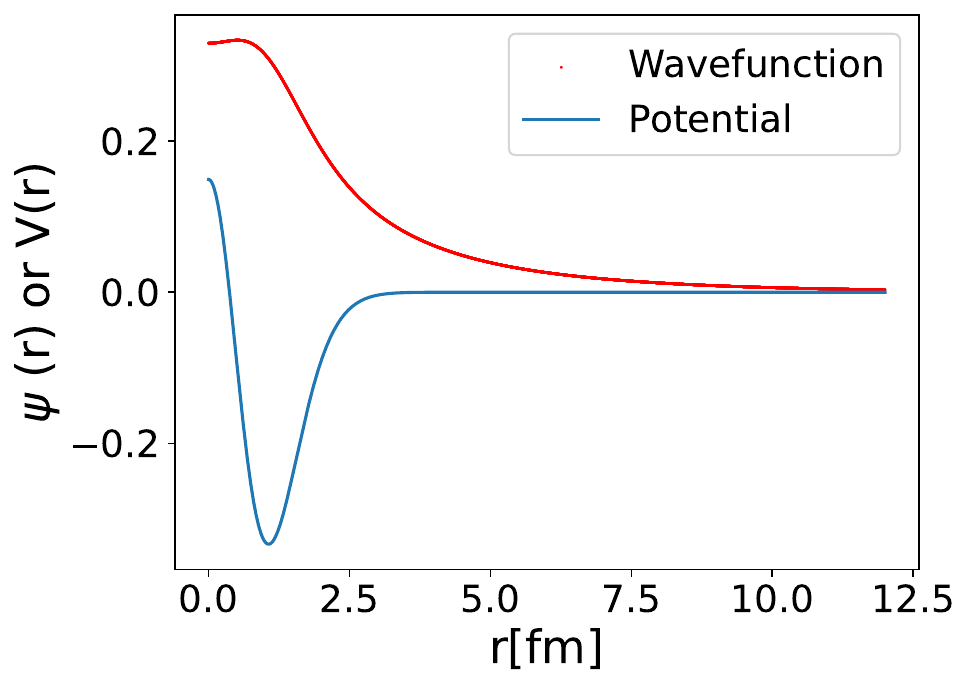}
    \caption{Ground state of the deuteron. The red dots show the predictions of the eigenfunction, as obtained after approximately 40000 training epochs. This is compared to the Minnesota potential in the $^3S_1$ channel as a blue line.}
    \label{Fig:plot_wf_r_Minn}
\end{figure}

\begin{figure*}[htbp]
    \centering
    \begin{subfigure}[b]{0.64\columnwidth}
        \includegraphics[width=\linewidth]{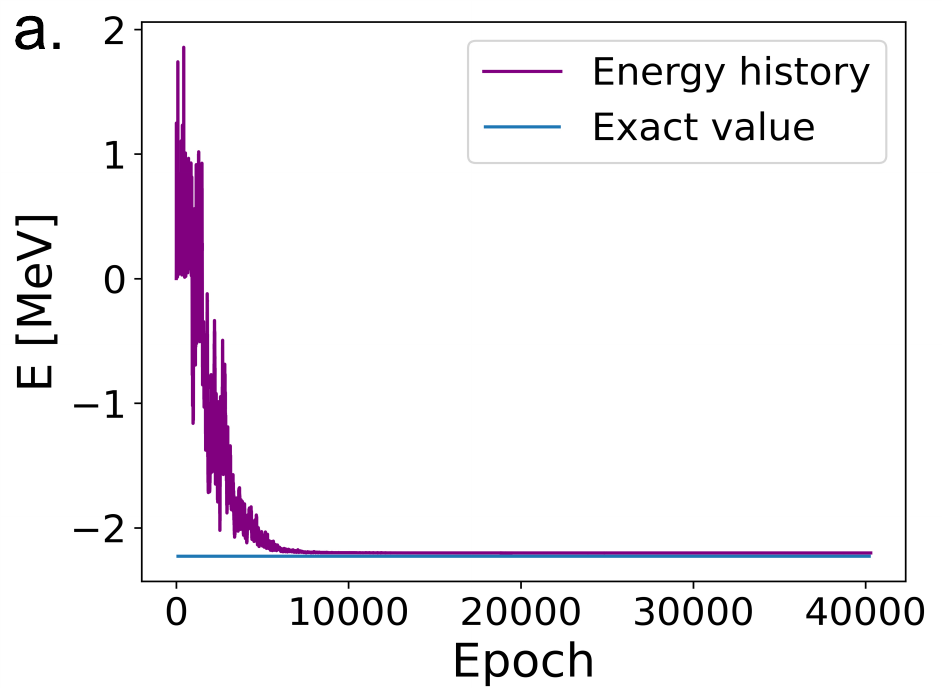}
    \end{subfigure} \hspace{0.2cm}
    \begin{subfigure}[b]{0.64\columnwidth}
        \includegraphics[width=\linewidth]{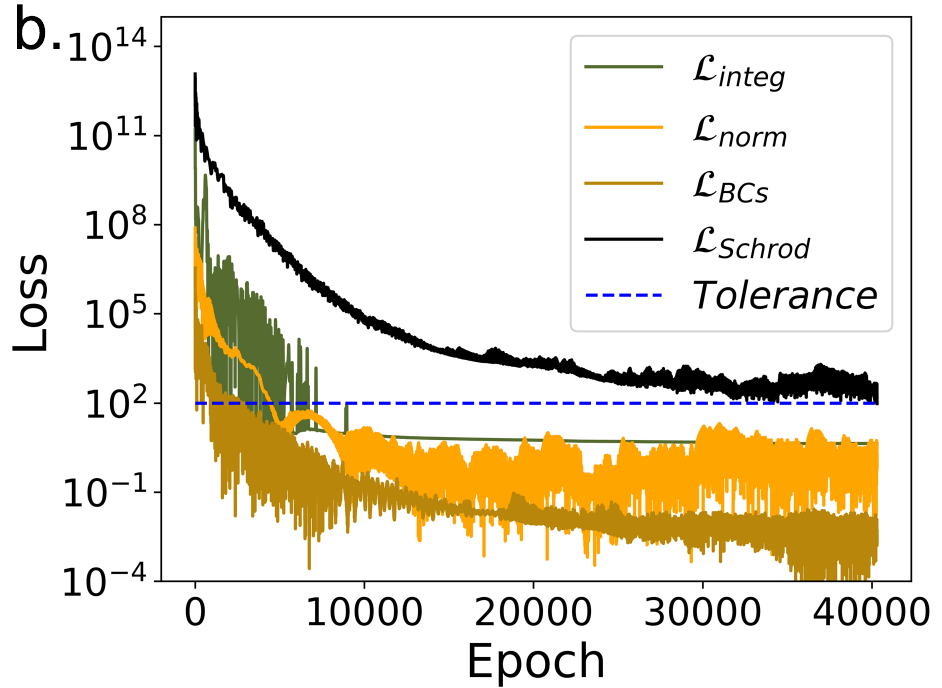}
    \end{subfigure} \hspace{0.2cm}
    \begin{subfigure}[b]{0.64\columnwidth}
        \includegraphics[width=\linewidth]{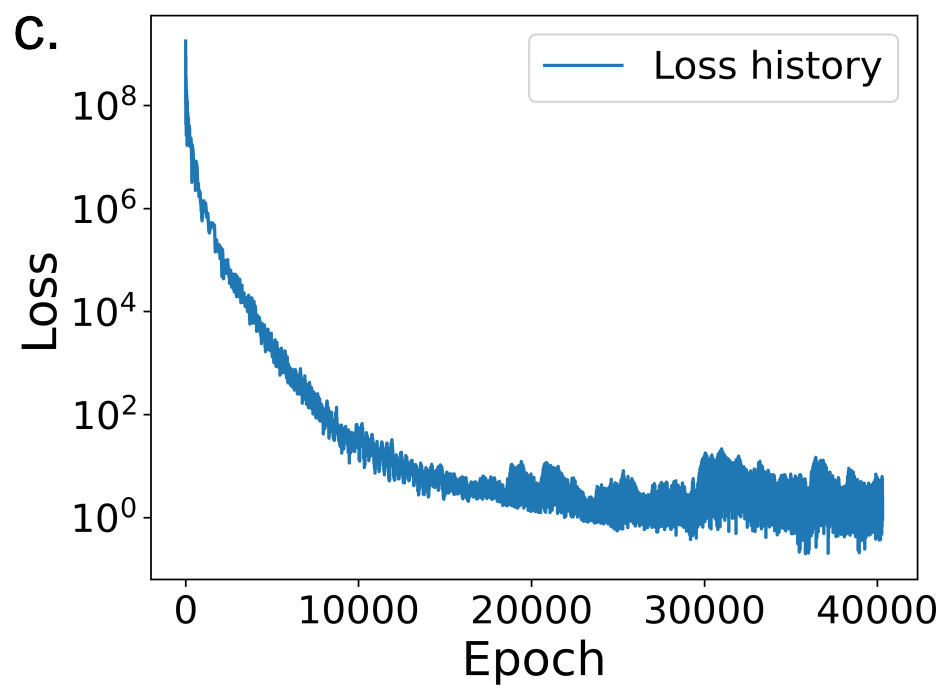}
    \end{subfigure}     
    \caption{Training history for the Minnesota potential. 
    (a) Behavior of the predicted nuclear eigenvalue. The purple line shows the energy of the deuteron predicted by the PINN, while the blue horizontal line is the experimental energy.
    (b) Partial contributions to the loss function. The green line is for the integral loss $\mathcal{L}_{integ}$, yellow is for the normalization loss $\mathcal{L}_{norm}$, brown shows the boundary conditions loss $\mathcal{L}_{BCs}$ and the black line is the differential equation loss $\mathcal{L}_{schrod}$. We considered a loss to be converged once it reaches below the dashed line, which corresponds to a hyperparameter that has been set to $100$. 
     (a) Total loss, Eq.~\eqref{eq: loss_overall}, through training.
     }
    \label{fig:train_r_Minn}
\end{figure*}

Since the Minnesota potential does not account for angular momentum mixing, the network had only one output for the auxiliary integral, Eq.~\eqref{eq: auxiliary_norm}, and one for the wavefunction in the $^3S_1$ spin triplet channel, $\psi_S(k)$.
The converged wavefunction is compared to the potential in Fig.~\ref{Fig:plot_wf_r_Minn} and follows the qualitative behavior of a bound $L$=0 wave, with an enhanced probability where the interaction is most attractive.
Figure~\ref{fig:train_r_Minn}a, shows the energy expectation value during training. It starts at a positive value and reaches near -2 MeV after 5000 epochs. Then, it gradually converges to a value close to the exact numerical solution with a final relative error of $1.1\%$.
Figure~\ref{fig:train_r_Minn}b demonstrates the behavior of each partial loss contributing to Eq.~\eqref{eq: loss_overall}. These are combined into the total loss shown in Fig.~\ref{fig:train_r_Minn}c through the weights from Table \ref{tab: weights_r}. We highlight that the differential equation loss starts converging later with respect to the other contributions shown in Figure~\ref{fig:train_r_Minn}b. This behavior has been induced intentionally by adjusting the~$\alpha_{Sch}$ weight, with the aim of making the network first learn the physical constraints of the system and later refine the solution of the differential equation. 

\subsection{Deuteron in momentum space} \label{sec: Res_k}

The $N{}^4LO$ and CD-Bonn interactions in momentum space are high precision models of the nuclear force and, therefore, they account for a small admixture of $D$ partial waves in the deuteron wavefunction. The mixing of angular momentum by the nuclear force is well known to be necessary for explaining the quadrupole moment of the deuteron. Consequently, the ground state wavefunction will have two components,
\begin{equation}
\label{eq: psi_deut}
    \psi(k) = \begin{pmatrix} \psi_S(k) \\ \psi_D(k) \end{pmatrix} \,,
\end{equation}
for the ${}^{3}S_1$ and ${}^{3}D_1$ partial waves. 
The Schr\"odinger equation for the deuteron reads

\begin{align}
\label{eq: schrod_deuteron}
E
\begin{pmatrix}
\psi_S(k) \\ \psi_D(k)
\end{pmatrix}
&=
\int_0^\infty
\begin{pmatrix}
V_{SS}(k, k') & V_{SD}(k,k') \\
V_{DS}(k,k') & V_{DD}(k,k')
\end{pmatrix}
\begin{pmatrix}
\psi_S(k') \\ \psi_D(k')
\end{pmatrix}
(k')^2 dk'
\nonumber\\
&\quad
+ \frac{k^2}{2\mu}
\begin{pmatrix}
\psi_S(k) \\ \psi_D(k)
\end{pmatrix}
\end{align}

where $V_{L\,L'}(k,k')$ is the interaction between partial waves $L$ and $L'$, $k$ is the momentum of the relative motion in MeV/c and 
\hbox{$\mu=m_p m_n/(m_p + m_n)$} is the reduced mass. We use the values $m_p=$~938.27208816~ MeV/c$^{\rm 2}$ and $m_n=$~939.56542052~MeV/c$^{\rm 2}$ for proton and neutron masses, respectively~\cite{CODATA2018}.
The normalization condition reads
\begin{equation}
    \label{eq: norm_psi_1}
    \mathcal{I}[\psi] = \int_0^\infty k^2\left(|\psi_S(k)|^2 + |\psi_D(k)|^2\right)\,dk = 1
\end{equation}
Note that $\psi_S(0)$ has finite value, while $\lim_{k\to0}\psi_D(k) = 0$.

\subsubsection{Network architecture and hyperparameters} \label{sec: net_k}

The neural networks we employ in momentum space is also of type feed forward, with $7$ hidden layers of $256$ neurons each and the momentum $k$ as the single input feature. Since we do not use the auxiliary output for normalization, the network will have only two outputs that are related to the wavefunction components of Eq.~\eqref{eq: psi_deut} by \hbox{$\phi_S(k) = k\,\psi_S(k)$} and \hbox{$\phi_D(k) = 10\, k\,\psi_D(k)$}. The factor $k$ in these definitions facilitates evaluating the $\mathcal{L}_{BCs}$ loss, as well as implementing Eq.~\eqref{eq: schrod_deuteron} into $\mathcal{L}_{Schrod}$. We have added a rescaling of a factor of 10 as a standardization correction for the $^3D_1$ wave, which gives a smaller contribution with respect to the $^3S_1$ counterpart.  With these definitions, Eq.~\eqref{eq: L_BC} is adapted to impose the boundary conditions 
    \begin{equation}
        \label{eq: bc_loss_deut}
        \phi_s(0) = \phi_d(0) = \phi_s(k_{max}) = \phi_d(k_{max}) = 0  \,,
    \end{equation}
where $M=k_{max}$ is the largest momentum point of an appropriate integration mesh, $\{(k_i,w_i): i=1,2,\ldots, N_c\}$, in k-space. This mesh covers the integration range $[0, k_{max} ]$, where we use $k_{max}=$ 1389.02~MeV/c with $N_c$=100 points for the $N{}^4LO$ and $k_{max}\approx$ 8573~MeV/c with $N_c$=200 points for CD-Bonn. Note that the mesh points are also used as collocation points throughout the simulations. We found that the training in momentum space is more efficient if we employ a finite difference method for the normalization. Thus, the integral~\eqref{eq: norm_psi_1} is evaluated as
\begin{align}
   \label{eq: norm_psi}
   \mathcal{I}[\psi] \approx{}& \sum_{i=0}^N w_i \,\left(|\phi_S(k_i)|^2 + \left|\frac{\phi_D(k_i)}{10}\right|^2\right) \,.
\end{align}
For the $\mathcal{L}_{var}$ loss, we implement the same hyperparameters used in coordinate space except for $E_{0} = -3$, $a = 0.01$. 
With all the above definitions, the overall loss for the PINN in momentum space is given by Eq.~\eqref{eq: loss_overall} but without $\mathcal{L}_{norm}$ and the last term in Eq.~\eqref{eq: norm_loss_deut}, which are specific to the auxiliary output method only.

\subsubsection{$N^4LO$ interaction}
\label{sec: N4lo}
\begin{figure}
    \centering
    \includegraphics[width=0.7\linewidth]{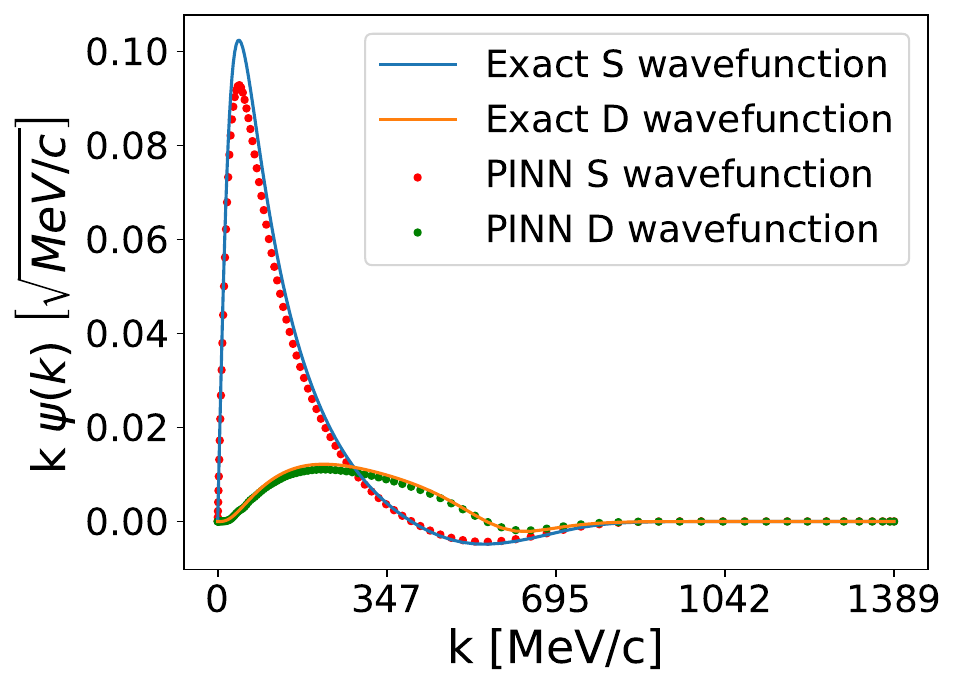}
    \caption{Ground state of the deuteron for the $N{}^4LO$ $\chi$EFT interaction. The red and green dots are the PINN solutions for the $^3S_1$ and $^3D_1$ wavefunction,  obtained after $4 \times 10^5$ epochs of training. The blue and orange lines are the exact solutions for the $^3S_1$ and $^3D_1$ components, respectively.}
    \label{fig: plots_deut_k_1}
\end{figure}

\begin{figure*}[tbp]
    \centering
    \begin{subfigure}[b]{0.64\columnwidth}
        \includegraphics[width=\linewidth]{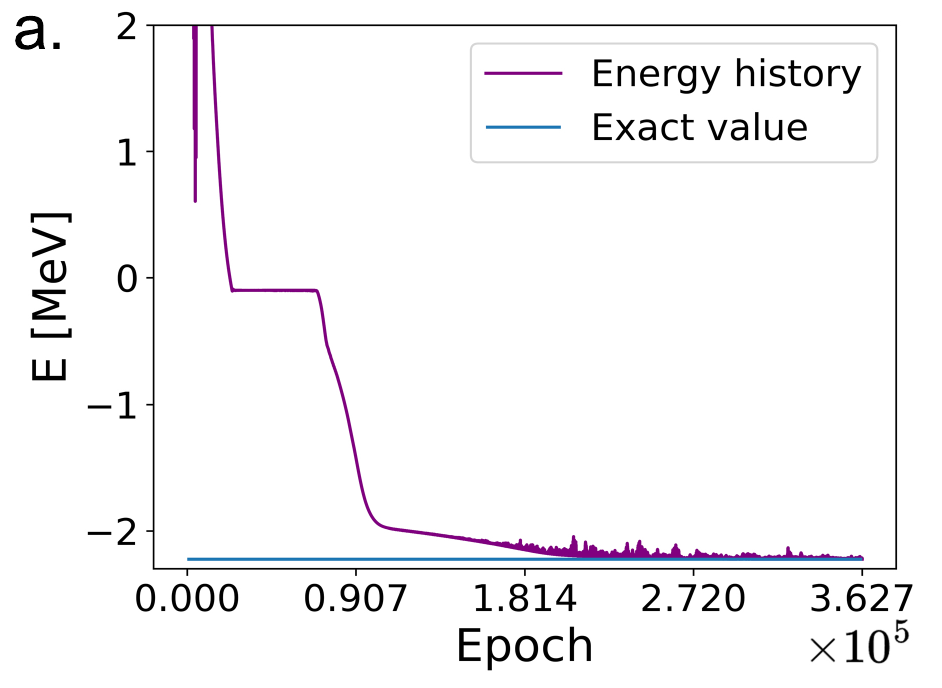} 
    \end{subfigure} \hspace{0.4cm}
    \begin{subfigure}[b]{0.64\columnwidth}
        \includegraphics[width=\linewidth]{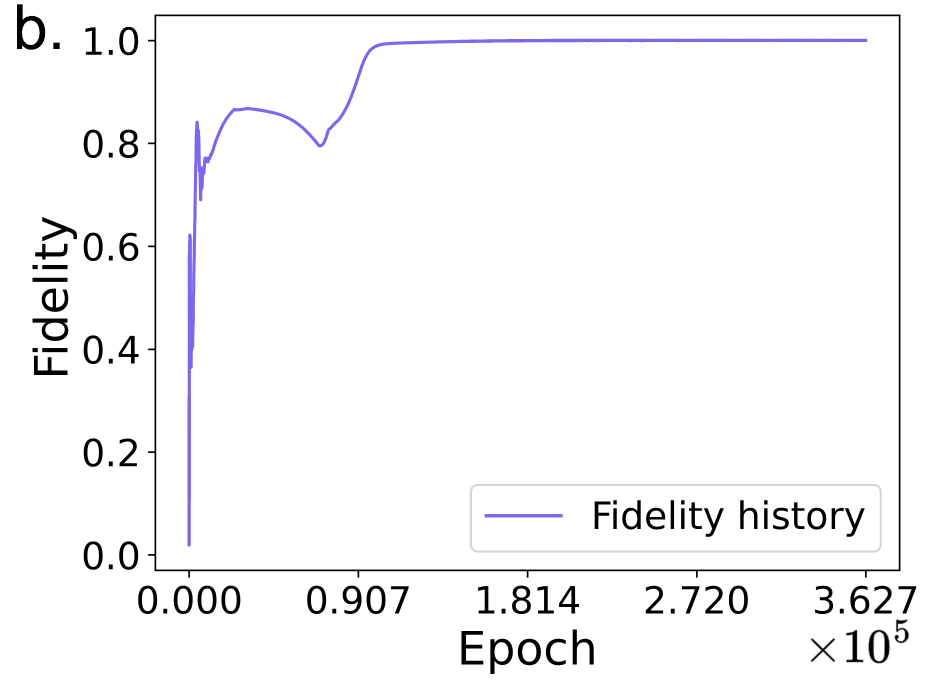} 
    \end{subfigure}  \hspace{0.4cm}
    \begin{subfigure}[b]{0.63\columnwidth}
        \includegraphics[width=\linewidth]{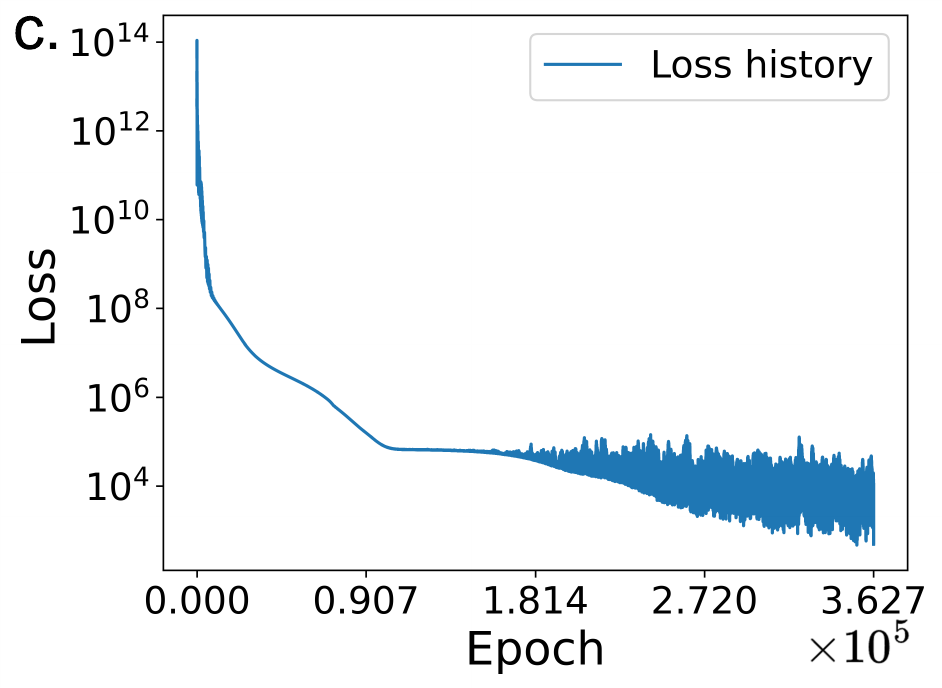}
    \end{subfigure} 
    \caption{Behavior of the PINN network for the $N^4LO$ interaction during training. (a) Ground state energy. The purple line shows the energy of the deuteron predicted by the PINN, while the blue horizontal line is the experimental energy. (b) Fidelity computed with respect to the exact diagonalization. (c) Total loss. }
    \label{fig: plots_deut_k_2}
\end{figure*}

With the 550~MeV/c cutoff of the $N^4LO$ interaction it is sufficient to exploit a uniform linear mesh mesh in the interval $[0,k_{max}]$ with $k_{max}\approx$~1400~MeV/c. This interval is sufficiently large to make the wavefunction vanishingly small at the boundaries but small enough to avoid instabilities from training the network on a too vast region where the both interaction and kinetic energy are negligible.
The network took around $4 \times 10^5$ epochs to reach convergence. The final relative error with respect to the exact numerical diagonalization of Eq.~\eqref{eq: schrod_deuteron} is $err_{E,Num}=-5.21\times10^{-5}$ with a fidelity of $ \mathcal{F}_\psi=0.9999933$, achieving more than acceptable levels of precision.

\begin{figure}
    \centering
    \includegraphics[width=0.7\linewidth]{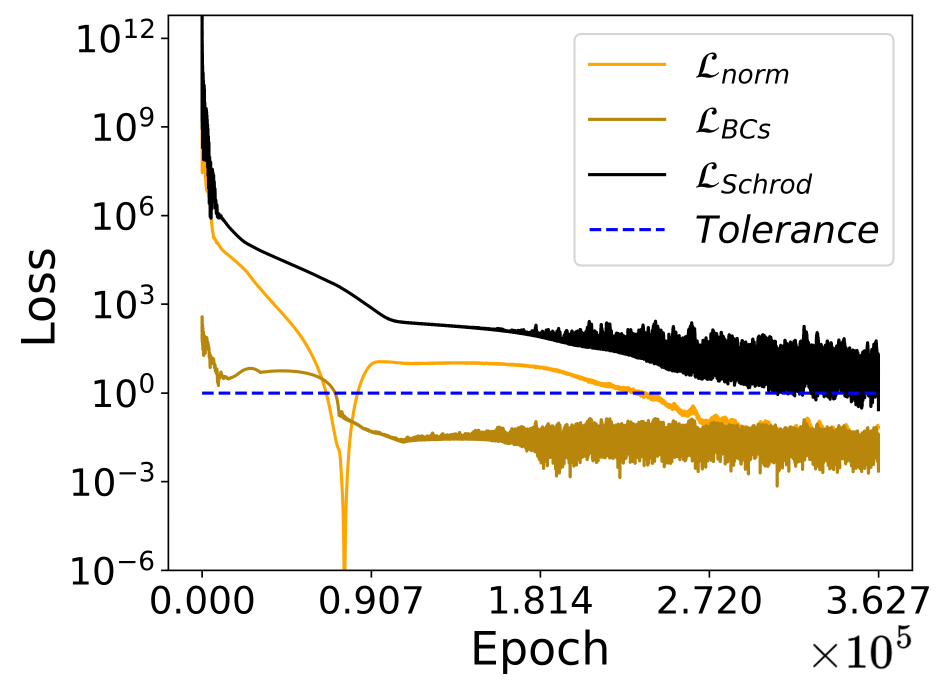}
    \caption{Behavior of the partial losses through training for the $N^4LO$ interactions. The yellow line is the normalization loss $\mathcal{L}_{norm}$, in brown is the boundary conditions loss $\mathcal{L}_{BCs}$ and the black line is the differential equation loss $\mathcal{L}_{Schrod}$. A loss is considered converged once it passes the blue dashed line, which is a hyperparameter, here set at $1$.}
    \label{fig: plots_deut_k_3}
\end{figure}

The PINN wavefunctions for both $^3S_1$ and $^3D_1$ components are compared to the exact diagonalization in Fig.~\ref{fig: plots_deut_k_1}. The curves agree very closely in shape, as it should be expected from the small above relative error and high fidelity in reducing the expected ground state energy of Eq.~\eqref{eq: E}. However, the PINN wavefunctions are slightly smaller indicating the the $\mathcal{L}_{norm}$ loss is not fully minimized and would require further training to improve although the energy is already converged. At the point were we stopped our training we have $\mathcal{I}[\psi]= 0.8219$.
Fig.~\ref{fig: plots_deut_k_1} also shows the difference in scale among the two components, with the $\psi_D(k)$ being significantly smaller. 

Figures~\ref{fig: plots_deut_k_2}a and \ref{fig: plots_deut_k_2}b demonstrate the trends of the energy and the fidelity during training. The energy increases rapidly in first few epochs and then it drops close to 0~MeV, which is the continuum threshold. It remains constant at this value for several iterations until around $8 \times10^4$ epochs, when it resumes dropping until it converges asymptotically. The behavior observed for the energy is reflected in the total loss in Figure \ref{fig: plots_deut_k_2}c. While overall the loss always trends downwards, the three distinct regions of Fig.~\ref{fig: plots_deut_k_2}a are reflected in different slopes for the overall loss. The rapid change of the energy in the first epochs correspond to a very steep descent of the loss but its slope is reduced during the epochs of constant energy. Eventually, the convergence of the energy is affected by an oscillating loss.  The cause of these behaviors can be understood by looking at the partial losses, in Figure \ref{fig: plots_deut_k_3}. All the losses quickly decrease in the first region. In the second region, $\mathcal{L}_{Schrod}$ decreases quite slowly, while $\mathcal{L}_{BCs}$ stagnates. The large value of $\mathcal{L}_{BCs}$ in this region indicates that the bound state boundary conditions are not met and scattering states are still admixed into the wavefunction. This is also reflected by the fidelity not being able to grow above 0.8. Once $\mathcal{L}_{BCs}$ drops below the tolerance threshold, the energy starts dropping below the continuum threshold (E$<$0~MeV) and the fidelity finally raises. This indicates that conditions of Eq.~\eqref{eq: bc_loss_deut} are finally met. 
The kink in $\mathcal{L}_{norm}$ around epoch $8 \times10^4$ signals that the normalization $\mathcal{I}[\psi]$ moves monotonically and it crosses the correct value of 1 at at the same epochs were the change in the structure of the wavefunction occurs. It then takes several other iterations before $\mathcal{L}_{norm}$ starts dropping properly. 
Note that the oscillations in the total loss are a consequence of noisy behavior of $\mathcal{L}_{BCs}$ and $\mathcal{L}_{Schrod}$ when they reach convergence.
This indicates a critical behavior in training the boundaries of the wave function.

Finally, we point out that the number of epochs required to train this neural network is one order of magnitude larger than the epochs needed to tune the PINN in coordinate space with the setup of Sect.~\ref{sec: res_r}. Despite this, each epochs of the $N^4LO$ training required around a tenth of computing time with respect to Sect.~\ref{sec: res_r} because of the different integration methods employed. The finite difference method means that PINN requires more epochs to reach convergence compared to the auxiliary output method. However, each epoch is significantly shorter because the automatic differentiation algorithm is executed only once per epoch, instead of four times as with the auxiliary output method.

\subsubsection{CD-Bonn interaction} \label{sec: cdbonn}

The CD-Bonn two-nucleon force is more demanding to diagonalize because it was specifically constructed to induce a very strong short-range repulsion~\cite{cdbonn}. This feature translates in small but non negligible components of the wave function at very high momenta. Correspondingly, we use a different mesh than for the $N^4LO$ with twice as many collocation points and that extends up to several $GeV/c$. We follow the standard approach of mapping a set of Gauss-Legendre quarature points into the [0,$\infty$) interval through a hyperbolic tangent transformation. This allows to extend to large momenta while keeping a dense mesh at low values of $k$. The largest momentum point in our mesh is $k_{max}\approx$~8573~MeV/c. 

We trained the solution of the CD-Bonn interaction by using the same neural network obtained for $N^4LO$---with its converged weights and biases---as the starting point. Since physics-informed neural networks are excellent interpolators, this ansatz provides a rough qualitative solution up to momenta of 1389~MeV/c where it was trained, greatly accelerating the training process. To avoid instabilities due to the initial extrapolation to larger momenta we first train the CB-Bonn interaction with the same momentum range used for the $N^4LO$ interaction and then gradually enlarge $k_{max}$ in steps of 400~MeV/c until we cover the full mesh at $k_{max}\approx$~8573~MeV/c.

\begin{figure}[tbp]
    \centering
    \includegraphics[width=0.7\linewidth]{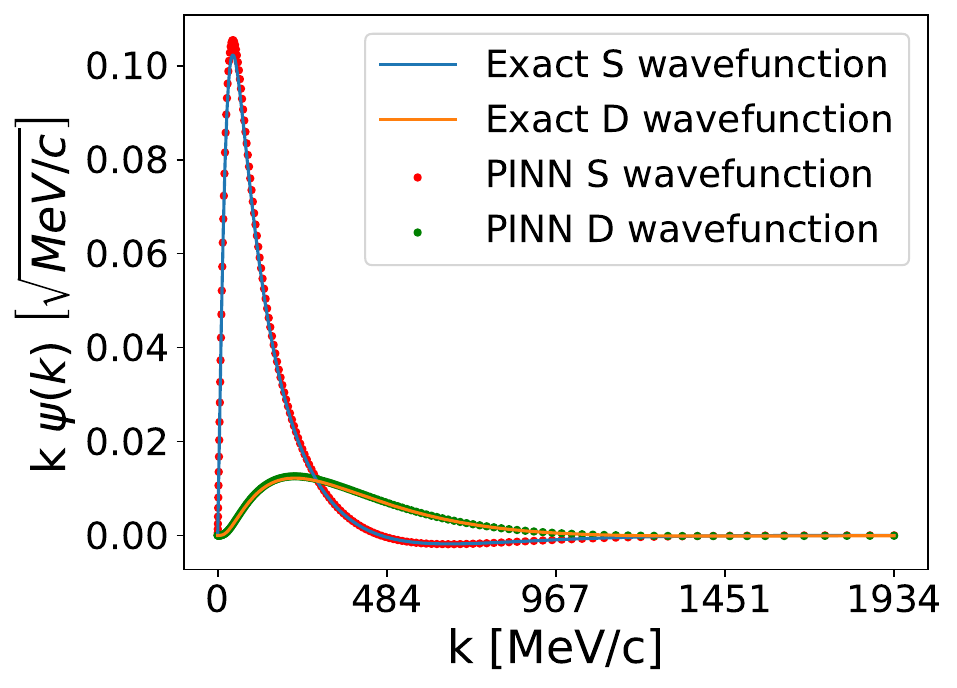}
    \caption{Ground state of the deuteron for the CD-Bonn potential. 
    The meaning of the curves is the same as for Fig.~\ref{fig: plots_deut_k_1}.
  }
    \label{fig: plot_cdbonn}
\end{figure}

Figure~\ref{fig: plot_cdbonn} demonstrated the final deuteron wavefunction components for momenta up to approx 2~GeV/c. In this case, the training strategy is sufficient to force the normalization condition so that the curves overlap properly. Note that the high-momentum tail would be visible only in logarithmic scale. The fidelity of the wavefunction with respect to the numerical benchmark is such that \hbox{$1-\mathcal{F}_\psi \approx 10^{-9}$}. The relative error for the deuteron binding energy with respect to the exact numerical benchmark is \hbox{$err_{E,Num} = -2.76\times10^{-7}$}, while it would be \hbox{$err_{E,Exp} = -6.01\times10^{-4}$} with respect to the experimental value. 

\subsection{Comparing the different interactions} \label{sec: compare}
The different final values we obtain for the relative errors with respect to the experimental energy mostly reflect how sophisticated is the interaction used. In fact, the error is maximum at $0.01$ for the very simple Minnesota potential, and minimum for the more modern $N^4LO$, at $-5.96\times10^{-6}$. We interpret this fact by observing  that the main source of error is from the potential itself and not the PINN method. In fact, when we compare the PINN results for $N^4LO$ to those for CD-Bonn, the older potential leads to worse results with respect to the experimental energy, with an error in the order of $10^{-4}$ compared to the $10^{-6}$ of $N^4LO$, while the error with respect to the numerical benchmark is slightly lower for CD-Bonn, probably due to the better start and the larger set of collocation points, at $-2.76\times10^{-7}$ compared to the $-5.76\times10^{-6}$ for $N^4LO$. However, the error for the Minnesota potential is still by far the highest even with respect to the numerical benchmark. This might reflect either a shortcoming in the auxiliary output method or just a need for longer training. Of course, all results will also be improved just by training for a longer period of time.

\section{Conclusions and future directions} \label{sec: conc}

We were able to utilize physics-informed networks to tackle realistic models of the nucleon-nucleon interaction in momentum space, even with strong high-momentum correlations, and to obtain highly accurate results on the deuteron  benchmarks. We also performed a similar simulation with a simplified interaction to prove the feasibility in coordinate space. The designed and validated neural network models are capable of computing the eigenvalue and eigenfunction of the ground state of the deuteron for these potentials.
The PINN framework is particularly interesting with respect to standard variational approaches because it is not specific to ground states but can be applied to the solution of excited states~\cite{Brevi2024}. In particular, the boundary condition loss can be naturally adapted to search for scattering solutions of the many-body problem.

The present implementation is specific to the one-dimensional radial Schr\"odinger equation and remains computationally more demanding than the direct exact diagonalizaton for the two-nucleon problem. Nevertheless, it demonstrates the feasibility of PINN for microsopic nuclear simulation and, in this sense, it paves the way for a completely new way to utilize novel machine learning tools for theoretical physics. 
To do so, it remains imperative to extend the present PINN implementation to full three dimensional space. For example, exploiting ans\"atze similar to those used in the VMC framework with standard variational and NQS wavefunctions. This will open the possibility of studying simple atomic nuclei and molecules. Further work in nuclear physics would also require the implementation of three-nucleon interactions.

\section*{Acknowledgements}
The Authors acknowledge the project CQES of the Italian Space Agency (ASI) for having partially supported
this research (Grant No. 2023-46-HH.0). The authors also acknowledge support from the Qxtreme project
funded via the Partenariato Esteso FAIR (grant No. J33C22002830006). 
AM acknowledges the IRA Programme, project no. FENG.02.01-IP.05-0006/23, financed by
the FENG program 2021-2027, Priority FENG.02, Measure FENG.02.01., with the support of
the FNP.
This work used the DiRAC Data Intensive service (DIaL3) at the University of Leicester, managed by the University of Leicester Research Computing Service on behalf of the STFC DiRAC HPC Facility (www.dirac.ac.uk). The DiRAC service at Leicester was funded by BEIS, UKRI and STFC capital funding and STFC operations grants. DiRAC is part of the UKRI Digital Research Infrastructure.


\bibliographystyle{elsarticle-num} 
\bibliography{refs.bib}

\end{document}